# Fast2Peak: High-Fidelity Restoration Algorithm by Single-Pixel Masking towards Real-Time Spectral Measurements

Nan-Nan Zhang[1,2#], Si-Heng Luo[3#], Xin-Di Pei[1#], Wen-Xuan Li[1], Jun Wang[1], Yi-Fan Bao[1], Hao Ma[1,2,4,*], Guo-Kun Liu[5,*], Bin Ren[1,2,4,*]

[1]State Key Laboratory of Physical Chemistry of Solid Surfaces, Collaborative Innovation Center of Chemistry for Energy Materials (iChEM), College of Chemistry and Chemical Engineering, Xiamen University, Xiamen 361005, China

[2]Institute of Artificial Intelligence, Xiamen University, Xiamen 361005, China

[3]Strait Institute of Flexible Electronics (SIFE, Future Technologies), Fujian Key Laboratory of Flexible Electronics, Fujian Normal University and Strait Laboratory of Flexible Electronics (SLoFE), Fuzhou, 350117, China

[4]Tan Kah Kee Innovation Laboratory, Xiamen 361005, China

[5]State Key Laboratory of Marine Environmental Science, Fujian Provincial Key Laboratory for Coastal Ecology and Environmental Studies, Center for Marine Environmental Chemistry & Toxicology, College of the Environment and Ecology, Xiamen University, Xiamen 361102, China

*Corresponding author: Hao Ma (oaham@xmu.edu.cn), Guo-Kun Liu (guokunliu@xmu.edu.cn), Bin Ren (bren@xmu.edu.cn)

## ABSTRACT

Improving the temporal and spatial resolution of spectroscopy inevitably reduces the photons available to each measurement, pushing rapid measurements into photon-starved conditions where weak spectral features are obscured by stochastic background noise. Here, we present Fast2Peak (F2P), a self-supervised Transformer framework that transforms the spectral restoration task into a context-prediction problem. Given that physically constrained spectral signals exhibit strong local continuity, whereas the stochastic noise fluctuates randomly and does not possess the continuity across neighboring spectral pixels, single-pixel masking enables F2P to restore the predictable signal structure while suppressing unpredictable fluctuations even without clean reference spectra. Unlike conventional offline spectral restoration, F2P achieves high-fidelity restoration within milliseconds, enabling real-time "acquire-and-process" operation. We demonstrate this capability in time-resolved electrochemical SERS and hyperspectral Raman imaging, where spectral restoration is performed during data acquisition. In addition, F2P has demonstrated its robust generalization across different instruments and spectral modalities, including Raman, NMR, XRD and FTIR.

## INTRODUCTION

There is a persistent drive to improve the temporal and spatial resolution of spectroscopic characterization techniques. High-speed acquisition or sampling from a small area or volume can reduce the number of detected photons, which pushes the measurement into photon-starved conditions where stochastic fluctuations dominate the observed signal and the signal-to-noise ratio (SNR) becomes extremely low[1–4]. This challenge becomes particularly high for Raman spectroscopy, [5–7]where Raman scattering is intrinsically weak. There are continuous efforts to shorten the acquisition time while retaining a high effective SNR.

However, restoring spectra acquired under photon-starved conditions, a task commonly referred to as spectral denoising, is an ill-posed inverse problem, and existing restoration approaches can only address parts of this challenge[8]. Conventional signal-processing and chemometric methods require case-specific parameter tuning and may attenuate weak or narrow vibrational features, leading to errors in quantitative analysis[9–14]. Supervised deep-learning methods can provide rapid and high-fidelity restoration after training, but generally rely on large-scale paired noisy-clean datasets or predefined noise models, neither of which is readily available across diverse experimental conditions[15–19]. More critically, the learned mappings are vulnerable to distribution shifts, which may distort spectra acquired from previously unseen samples, instruments, or noise regimes[19]. Self-supervised learning (SSL) methods do not require clean reference spectra, but often need spectrum-specific optimization or have to compromise fine spectral details to achieve computational efficiency, limiting their use in real-time measurements[20–24]. These limitations call for an automatic, real-time, and high-fidelity spectral restoration method that can be generalized across instruments, samples, and acquisition conditions, even in extreme photon-starved regimes.

A key challenge in self-supervised spectral restoration is therefore to distinguish the underlying spectral feature from stochastic noise. It is important to recognize that spectral features, including peak positions, widths, intensities, and local line shapes, exhibit strong contextual correlations along the spectral axis, whereas stochastic detector noise and photon noise are assumed to be statistically independent across the neighboring spectral pixels (Figure 1a). Therefore, the different properties of signal to noise provide a simple but powerful route to self-supervised restoration: if some discrete spectral pixels are hidden, their signal values can be inferred from their adjacent context, whereas independent noise cannot be reliably predicted. With this understanding, we develop Fast2Peak (F2P), a self-supervised Transformer framework that masks individual spectral pixels and forces their restoration from the remaining spectral context. F2P is able to learn spectral structure while suppressing unpredictable fluctuations. We start with Raman spectroscopy, which is characterized by low photon counts and sharp, overlapping peaks and serves as an exceptionally demanding modality for spectral restoration. We demonstrate F2P's strong restoration performance across diverse Raman peak profiles and instruments, as well as its applicability to other spectral modalities, including NMR, XRD, and FTIR. F2P further decouples pretraining from inference, thereby avoiding spectrum-specific optimization and enabling high-fidelity restoration with millisecond-scale inference. When integrated into the software of an instrument, F2P transforms the conventional "acquire-then-process" measurement into a real-time "acquire-and-process" operation, as demonstrated in time-resolved electrochemical SERS and hyperspectral Raman imaging.

# EXPERIMENTS AND ALGORITHMS

**Statistical difference between signal and noise.** As illustrated in Figure 1a, spectral peaks in a clean spectrum $x$ extend across neighboring spectral pixels because of their finite linewidths, which are determined by intrinsic broadening mechanisms and instrumental resolution. The signal values at neighboring pixels $p^i$ and $p^j$ are therefore correlated. In contrast, stochastic detector or photon fluctuations ($n$) are assumed to be statistically independent between neighboring pixels. With this understanding, the underlying signal intensity at a masked pixel can be estimated from the neighboring spectral context, whereas the random noise fluctuation at that pixel cannot be reliably predicted. F2P uses this difference to learn predictable spectral features while suppressing stochastic fluctuations, without requiring clean reference spectra for training (The detailed derivation is provided in the Supporting Information).

**Single Pixel Masking**. Based on this statistical difference, we designed a single-pixel masking strategy. For an input spectrum $y$, spectral pixels are independently and randomly selected for masking (Figure 1b). At each selected position $p^k$, the original intensity value is replaced by a value randomly sampled from a local window size of $\gamma$ ($\gamma = 7$), spanning $p^{k-3}$ to $p^{k+3}$. The replaced intensity is required to be different from the original intensity at $p^k$. This procedure produces the masked spectrum $y_m$ and removes the original intensity from each selected input position. This single-pixel replacement ensures that the network will not simply "copy" the target intensity from the input. Instead, the model is encouraged to restore the clean signal at these masked positions by capturing the inherent contextual correlations within the remaining spectral features of $y_m$.

**Model Training.** Figure 1c illustrates the overall self-supervised training pipeline of F2P. The simulated training set comprises a total of 400,000 spectra, as described in the Dataset section of the Supporting Information. For each spectrum $y$ in the training set, single-pixel masking (Figure 1b) is first applied to obtain $y_m$. The masked spectrum $y_m$ is then fed into the Spectral Transformer for self-supervised learning to obtain the restored spectrum $f_\theta(y_m)$, where $\boldsymbol{\theta}$ denotes the trainable parameters. These parameters are optimized by minimizing the Mean Squared Error (MSE) loss, formulated as:

$$\theta = \arg\min_{\theta} \mathbb{E} \sum_{i} \left[ \left\| [f_{\theta}(y_m)]_i - y_i \right\|_2^2 \right] \quad (1)$$

where $y_i$ and $[f_\theta(y_m)]_i$ denote the values of the training spectrum and the masked spectrum at the masked position $i$, respectively. At each epoch, the $f_\theta(y_m)$ is compared against the original spectrum $y$ only at the masked locations.

We trained F2P in PyTorch on an NVIDIA RTX 3090 GPU using the AdamW optimizer for 1,000 epochs. The initial learning rate was 0.0001 and was reduced through cosine annealing to one-tenth of its initial value by the end of training.

**Model Inference.** As illustrated in Figure 1d, to ensure structural compatibility across different instruments with varying spectral resolutions and input lengths, bicubic interpolation is employed to resample all input spectra to a standardized length of 1600. The resampled spectrum can then be directly processed by the pre-trained model without additional optimization. Following the restoration process, the output signal is rescaled to its original spectral length through a post-processing interpolation step to maintain the physical correspondence of the frequency axis. The architecture of the Spectral Transformer used for training and inference is detailed in Figure S1.

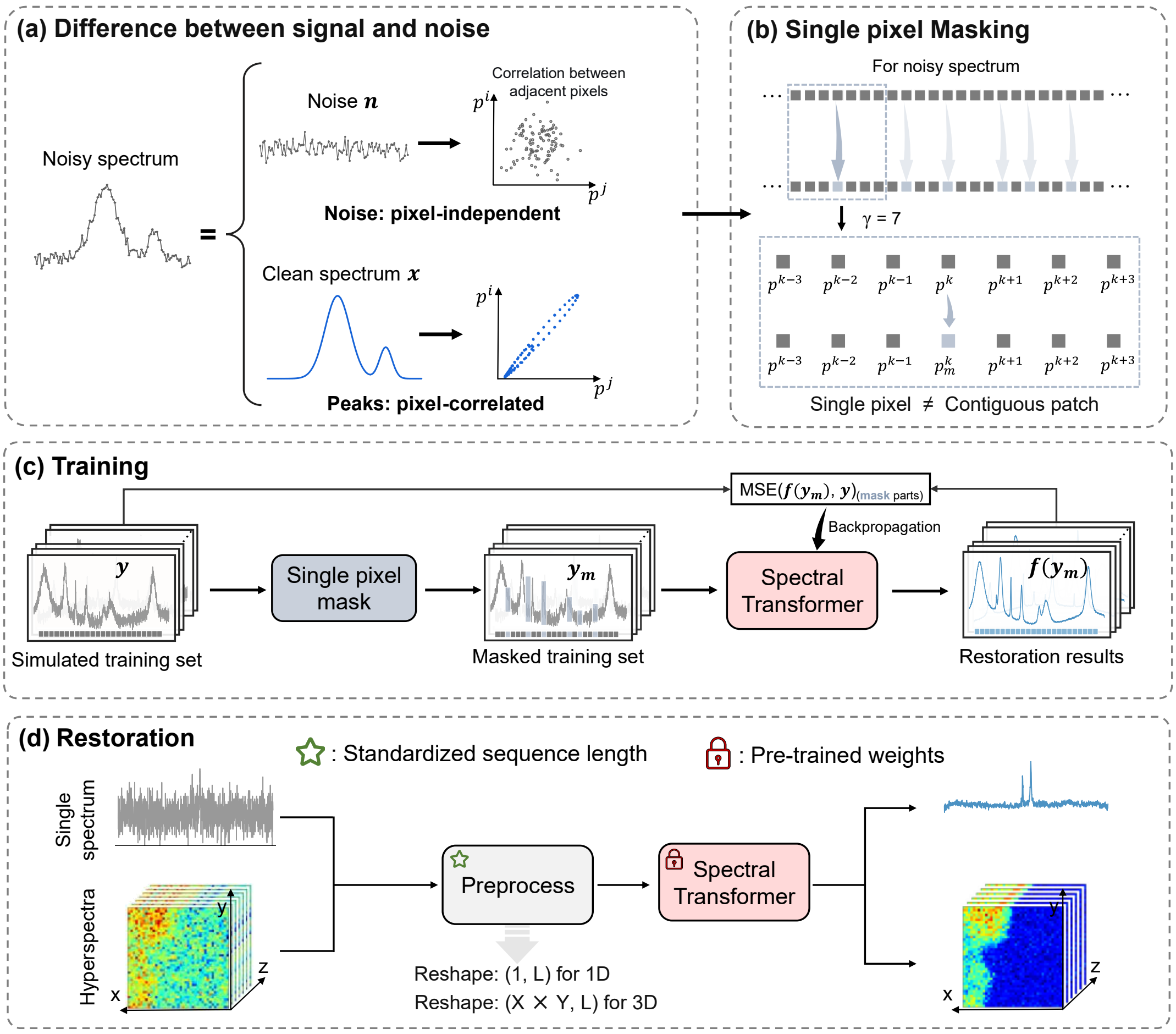

**Figure 1.** Principle and workflow of F2P for self-supervised spectral restoration. (a) Statistical difference between signal and noise, (b) single-pixel masking strategy, (c) the overall self-supervised training process, and (d) restoration of single spectra/hyperspectral data using the pre-trained Spectral Transformer.

## RESULTS AND DISCUSSION

### High-Fidelity Spectral Restoration

To rigorously evaluate the performance of denoising algorithms across diverse chemical signatures of Raman spectrum, we selected a representative spectrum from the simulated test set (Simu set) of the F2P Benchmark (F2P-B; details of dataset construction are provided in the Supporting Information). The reference Raman spectrum contains two strong peaks at 300 $cm^{-1}$ and 900 $cm^{-1}$, a weak peak at 450 $cm^{-1}$, a sharp peak at 775 $cm^{-1}$, and overlapping bands in the 580-675 $cm^{-1}$ region (black traces, Figure 2a). We added different levels of noise to obtain five noisy spectra (gray traces, Figure 2a). The five noisy spectra had pSNR values of 4.0, 2.5, 1.5, 1.3, and 1.0, respectively, calculated using the weak peak at 450 $cm^{-1}$. The definitions and calculation formulas for pSNR and other evaluation metrics are provided in the Supporting Information.

Notably, different restoration algorithms exhibit varying capabilities in processing these distinct spectral profiles. At a high raw-spectrum pSNR ($pSNR_{RAW}$) of 4.0 (left of Figure 2a), no obvious performance difference is observed among the tested methods. As the $pSNR_{RAW}$ gradually declines to 2.5 and 1.5, a degradation in restoration fidelity is observed across all tested methods. The traditional filters (SG, WTD and PEER) and earlier deep learning methods (ResUNet, SMAE and P2P) begin to show obvious distortions. For instance, the weak peak at 450 $cm^{-1}$ is frequently submerged by residual artifacts. In contrast, the sharp peak at 775 $cm^{-1}$ suffers from excessive broadening and even the local root mean square error (RMSE) paradoxically exceeds that of the raw spectrum (Figure 2b), indicating a failure to preserve peak morphology. F2P maintained the lowest RMSE (0.0117) for weak peak at 450 $cm^{-1}$ and preserved sharp structures far better than SMAE (Figure 2b, $pSNR_{RAW}$=1.5), demonstrating superior resilience by maintaining high-fidelity restoration across all peak profiles.

Restoration became more difficult when $pSNR_{RAW}$ decreased to 1.3 and 1.0. At $pSNR_{RAW}$ = 1.0, the weak peak at 450 $cm^{-1}$ was no longer clearly distinguishable from noise in the raw spectrum (Figure 2a, right panel). Under these conditions, assessing restoration requires comparison with the reference spectrum, because suppressing fluctuations alone does not indicate the preservation of the underlying peak. The overlapping bands in the 580-675 $cm^{-1}$ region (two peaks centered at approximately 605.25 and 640.28 $cm^{-1}$) provide a further test of peak preservation. Their full widths at half maximum are 50 and 70 spectral pixels, respectively, and their peak separation is 60 pixels. The comparable peak widths and separation produce substantial overlap, while individual $pSNR_{RAW}$ values as low as 1.48 and 1.55 make the two peak profiles difficult to distinguish from noise. Under the same noise conditions, F2P preserved these overlapping profiles more faithfully than the other tested methods (Figure 2). Together, these comparisons show that F2P better preserves the weak, narrow, and overlapping peaks in this representative spectrum.

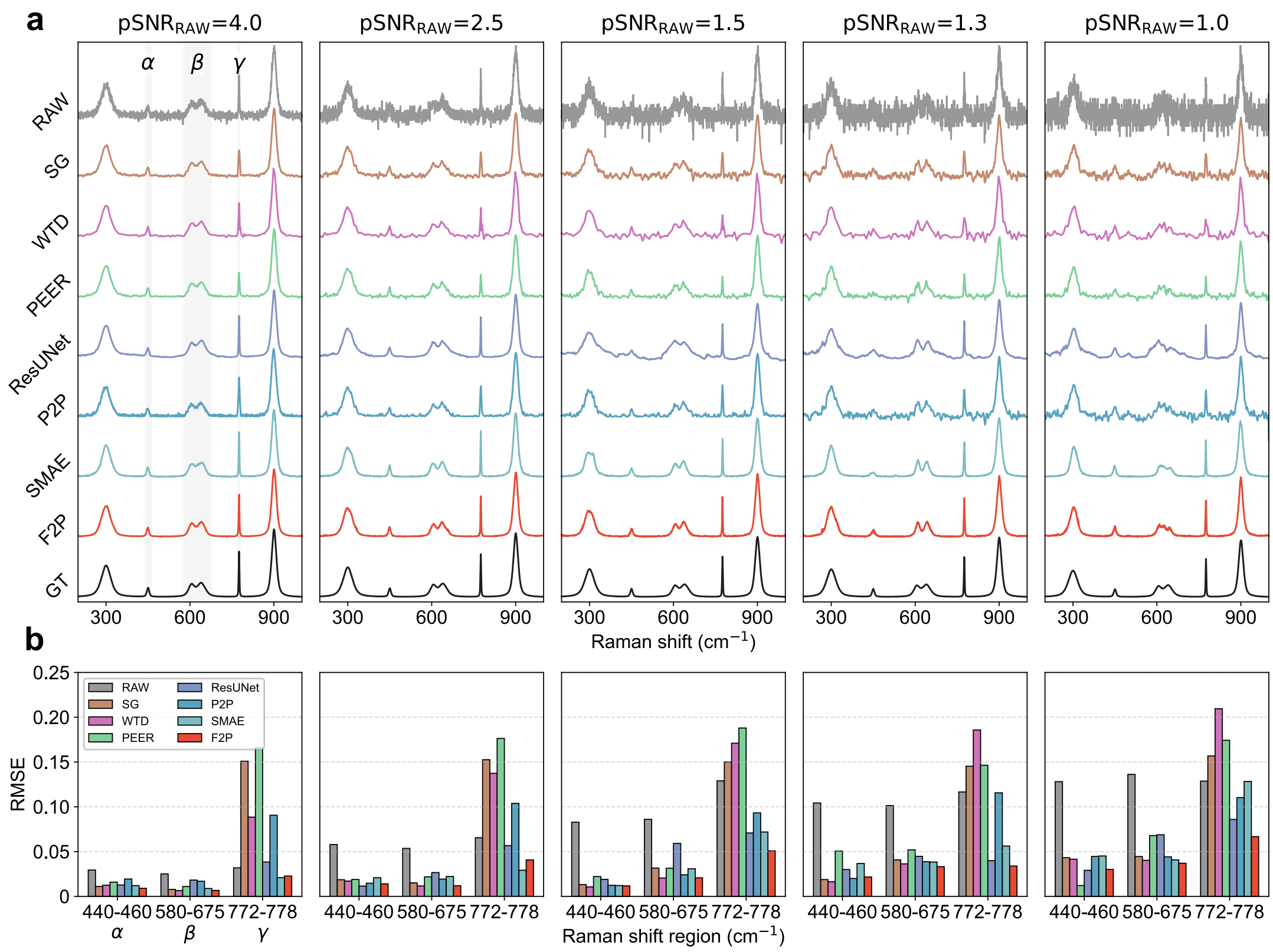

**Figure 2**. (a) Comparison of F2P with competing algorithms for denoising spectra with different peak shapes and pSNR (4.0, 2.5, 1.5, 1.3 and 1.0). (b) RMSE of fidelity for different peak shapes with different methods.

To further demonstrate the performance of F2P beyond individual examples, we systematically evaluated its performance on all 1024 spectra in the Simu test dataset of F2P-B, with a mean raw SNR of 17.95 dB. As summarized in Table 1, F2P demonstrated state-of-the-art performance with a SNR of 33.62 dB and the lowest RMSE of 0.0079, among the methods evaluated. Traditional methods (SG, WTD and PEER) improved SNR, but were limited by their complex parameter tuning. Supervised ResUNet outperformed these traditional approaches by leveraging clean reference spectra for explicit target guidance. The recently reported self-supervised methods (P2P and SMAE) underperformed ResUNet, owing to the constrained training strategy and the lack of external references. Notably, through large-scale self-supervised pretraining, F2P surpassed the supervised ResUNet, achieving a 3.59 dB higher SNR and approximately 32% lower RMSE without requiring clean reference spectra. Furthermore, the pretrained F2P model can be directly applied to unseen spectra without spectrum-specific optimization, enabling substantially higher computational efficiency than other self-supervised methods. F2P processed all 1,024 spectra in only 0.3 s, approximately 11 times faster than SMAE and over four orders of magnitude faster than P2P. These results strongly suggest that F2P can simultaneously extract low-abundance molecular signatures and preserve fine structural details with the highest processing speed.

**Table 1.** Evaluation results of F2P on the Simu test dataset.

| Methods | SNR (dB) ↑ | RMSE ↓ | Total Time (s) ↓ |
| --- | --- | --- | --- |
| Raw | 17.95 ± 8.73 | 0.0505 ± 0.0286 | - |
| SG-23-3 | 25.49 ± 8.03 | 0.0191 ± 0.0089 | 0.5041 ± 0.048 |
| WTD-4 | 26.24 ± 8.34 | 0.0180 ± 0.0094 | 0.7366 ± 0.0529 |
| PEER | 26.09 ± 8.02 | 0.0181 ± 0.0089 | 144.59 ± 0.4406 |
| ResUNet | 30.03 ± 6.66 | 0.0117 ± 0.0067 | 0.4783 ± 0.003 |
| P2P | 26.63 ± 7.43 | 0.0171 ± 0.0096 | ~ 20000.00 |
| SMAE | 29.07 ± 5.88 | 0.0119 ± 0.0054 | 3.4452 ± 0.0189 |
| **F2P** | **33.62 ± 7.74** | **0.0079 ± 0.0045** | **0.3080 ± 0.004** |

**Advantages of Single-pixel masking**

In contrast to conventional patch-based strategies, F2P employs a single-pixel masking mechanism (Figure 1b), transforming denoising into a task of recovering signal distributions from the global context. This allows the Transformer to use the correlations between neighboring spectral points to reconstruct the masked intensity while suppressing random fluctuations. As shown in Figure S2a, F2P achieves the best restoration performance when using a random single-pixel masking with RMSE of 0.0079 (Table 1). As the masking scale increases from 2 to 16 pixels, the restoration accuracy declines significantly, indicating that larger masking scales increasingly distort the local spectral context required for precise restoration. Consistently, removing the masking module eliminates the restoration capability of F2P, leading to degraded performance as the model tends to behave as an identity mapping that retains high-frequency noise (Figure S3). This sensitivityto mask size suggests that the single-pixel mechanism may be generalized beyond F2P. We incorporated this mechanism to SMAE, a representative patch-based masked autoencoder originally designed for global representation learning for classification. As illustrated in Figure S2b, reducing patch size of SMAE down to 1 significantly improves its denoising performance with RMSE decreasing from 0.0988 to 0.0119. Nevertheless, downsampling in SMAE limits the preservation of fine-grained contextual information, and its restoration error remains higher than that of F2P.

The restoration error depends on the masking ratio in a U-shaped way with optimal performance at approximately 10% (Figure S2c). It indicates that excessive masking removes essential local context, whereas insufficient masking weakens the denoising task and increases the risk of directly reproducing the input spectrum. Sensitivity analysis of replacement range $\gamma$ (Table S3) only shows $< 3\%$ RMSE fluctuation, indicating that the restoration performance mainly depends on the surrounding spectral context rather than the masked intensity.

**Noise-agnostic Generalization across Instruments and Spectral Modalities**

Although F2P has achieved near-state-of-the-art performance across diverse experimental systems (Figure S4), the inherent physical differences in optical alignment and detector (CCD/CMOS) among different Raman spectrometers may produce very different noise patterns. We thereby evaluated the performance of F2P across instruments.

Figures 3a-e depict the Raman spectra of SSEs (solid electrolyte datasets of F2P-B) from five commercial spectrometers, i.e., LabRAM HR Evolution (HR), Raman-11, InVia Raman, Xplora,

and i-Raman. F2P consistently achieves high restoration performance, regardless of the vast differences in SNR and background of raw spectra across different instruments. As shown in Figure 3a, for the high-resolution HR system, F2P significantly enhances spectral smoothness while strictly preserving sharp vibrational features across all integration times. For the other research-grade systems, including Raman-11, Invia, and Xplora (Figure 3b-d), F2P demonstrates robust adaptability to varying SNRs and backgrounds. Notably, F2P precisely restores critical spectral features even when they are almost entirely buried. For instance, for the case of the 30 ms integration time in the Raman-11 system (Figure 3b), critical features of the $\tau_0(CH_2)$ at 1235 $cm^{-1}$, $v_s(CF_3)$ at 1247 $cm^{-1}$ and the peak of δ(SNS) at 740 $cm^{-1}$, can be clearly restored.[25] The performance on Raman spectra obtained by Xplora further demonstrates that F2P can accommodate large baseline variations, reducing high-frequency noise while preserving the broad baseline. Meanwhile, for the portable i-Raman system (Figure 3e), which inherently possesses lower spectral resolution, F2P effectively reduces noise while preserving the broader spectral profiles. Together, these results show that F2P substantially mitigates instrument-dependent variations and provides robust Raman denoising across various commercial spectrometers.

F2P also demonstrated zero-shot denoising of infrared (IR), nuclear magnetic resonance (NMR), and X-ray diffraction (XRD) datasets (Table S4-6) without any modality-specific fine-tuning using the pretrained model directly. From an information-restoration perspective, this wide generalization is rooted in a fundamental principle shared by diverse spectral measurements: physically meaningful features possess intrinsic local correlations, whereas stochastic noise is inherently unpredictable from neighboring pixels. This explains why structured signatures from diverse modalities (vibrational, magnetic resonance, or diffraction features) can be reliably restored even when they are embedded in disparate noise environments such as shot, thermal, or electronic noise. By leveraging a single-pixel masking strategy to isolate these independent noise fluctuations, F2P learns the continuous spectral structure from neighboring information while suppressing unpredictable noise, enabling the pretrained model to be directly applied to different spectral modalities without additional training.

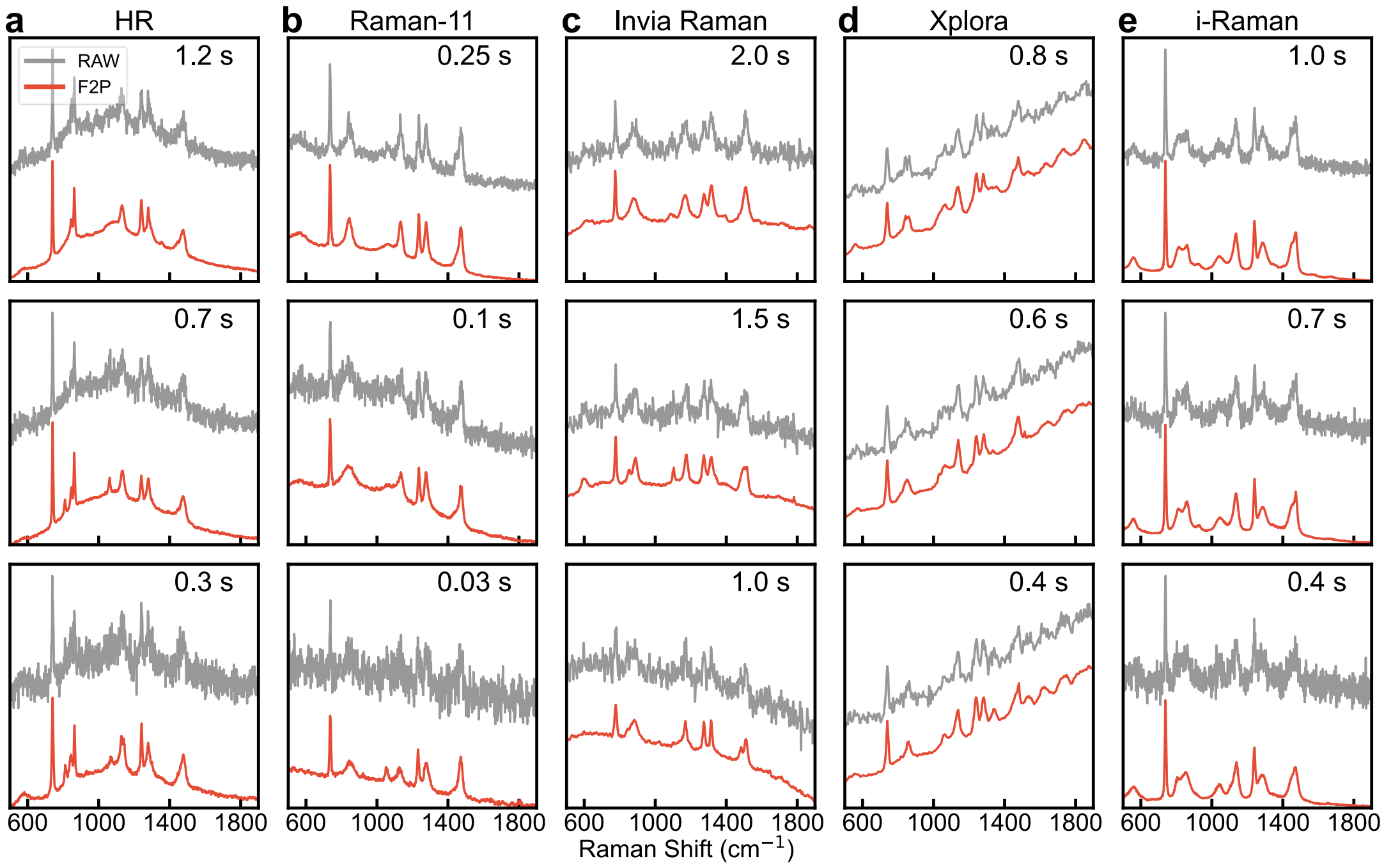

**Figure 3.** Generalizability validation of F2P on experimental spectra of the electrolyte PEO + LiTFSI collected on different instruments: (a) HR evolution (Horiba); (b) Raman-11(Nanophoton); (c) Invia Raman (Renishaw); (d) Xplora (Horiba); (e) i-Raman (B&W TeK).

**Towards an Acquire-and-process paradigm**

Conventional denoising workflows inevitably impose a "dark period" of offline post-processing, delaying the feedback loop between measurement and decision-making. This latency is particularly detrimental for experiments demanding real-time spectral fidelity: transient interfacial dynamics (where reaction intermediates evolve faster than post-processing) and rapid hyperspectral chemical mapping (where spatial throughput is limited by computational bottlenecks). By compressing this latency to the millisecond scale, F2P embeds spectral reconstruction directly into the acquisition pipeline, potentially transforming reacting data collection into active, closed-loop experimentation.

We evaluated F2P using time-resolved electrochemical surface-enhanced Raman spectroscopy (EC-SERS) of Nile Blue adsorbed on an Ag nanoparticle-modified Ag electrode (Figure 4a). Nile Blue undergoes a reversible potential-driven transformation from the non-resonant reduced state ($NB_{red}$) to the resonant oxidized state ($NB_{ox}$). The potential-dependent appearance of $NB_{ox}$ bands provides a spectroscopic indicator of this interfacial redox conversion. Measurements were performed using an EC-Raman-L215 electrochemical Raman microscope (Jayitec, Xiamen, China). Raman spectra were acquired with 20 ms integrations during cyclic voltammetry between -0.20 and +0.30 V at 1 V $s^{-1}$. A simulated TTL trigger synchronized the spectrometer with the embeded potentiostat, allowing each spectrum to be associated with a nominal electrode potential. Each integration spanned 20 mV on the linear portions of the sweep, defining the potential interval sampled by each spectrum.

The raw spectra acquired with 20 ms integrations contained substantial noise, which obscured changes in the characteristic bands during potential cycling. Following F2P restoration, the cyclic variation of the Raman bands became clearer in the Raman maps (Figure 4b). Representative spectra at nominal potentials of 0.160, 0.180 and 0.200 V also showed a more readily discernible potential-dependent change in the band near 594 $cm^{-1}$ after restoration (Figure 4c). Together, these observations show that F2P improves the visibility of redox-sensitive spectral features in EC-SERS measurements acquired with 20 ms integrations, with the 100 ms measurements serving as a reference (Figure S5), helping to follow the molecular response during rapid potential cycling.

Beyond EC-SERS, F2P was applied to offline $WSe_2$ hyperspectral Raman imaging (Figures S6 and S7) and integrated into an asynchronous end-to-cloud pipeline for real-time graphene imaging (Figure S8). Under a 50 ms integration time, F2P progressively restored clear graphene boundaries and characteristic Raman features during scanning, as shown in Videos S1. Quantitatively, the spatial resolution improved from 4.57 to 1.24 μm, corresponding to an approximately 3.7-fold enhancement (Table S7).

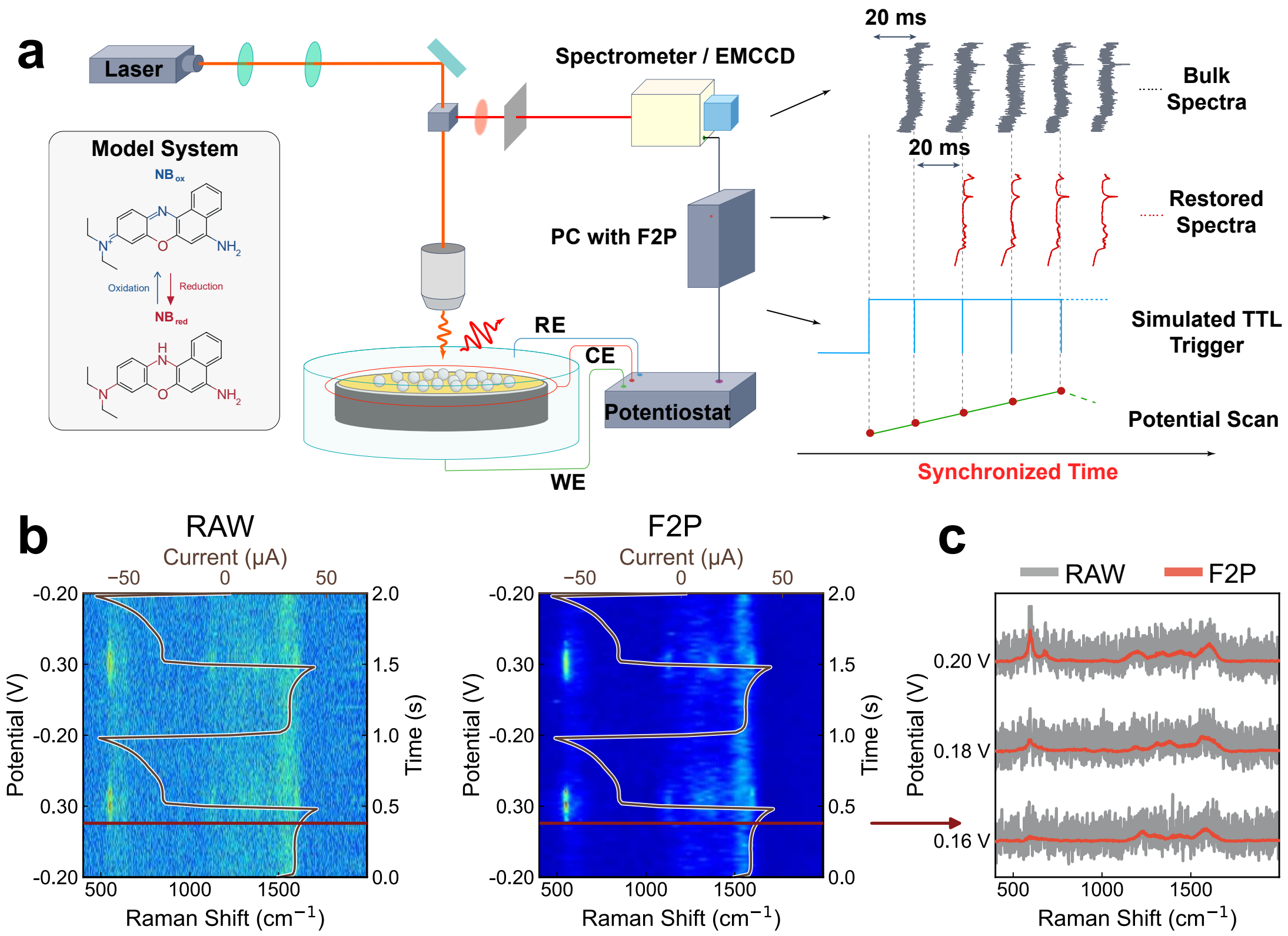

**Figure 4. F2P restoration of time-resolved EC-SERS detection of Nile blue.** (a) Synchronized TEC-SERS-F2P acquisition workflow using a simulated TTL trigger. (b) Raw and F2P-restored potential-resolved Raman maps acquired with 20 ms integration time, with the measured current overlaid. (c) Representative raw and F2P-restored spectra near 0.18 V.

# CONCLUSION

In summary, we proposed Fast2Peak (F2P), a self-supervised Transformer-based framework that advances the spectral restoration in the photon-starved regime. By leveraging the distinct statistical properties of physically constrained spectral signals and stochastic noise, F2P achieves state-of-the-art restoration fidelity without requiring any clean ground-truth labels. Based on the newly established F2P-B benchmark, F2P significantly outperforms existing self-supervised and traditional methods, also demonstrating exceptional generalization capabilities across diverse chemical systems, varied commercial instruments, and distinct spectral modalities. More importantly, by decoupling pre-training from forward inference, F2P enables a paradigm shift from conventional “acquire-then-process” to a real-time “acquire-and-process” workflow, and has been successfully applied to time-resolved EC-SERS and hyperspectral Raman imaging. This high-throughput, low-latency infrastructure provides a robust foundation for the next generation of AI-driven autonomous material characterization and high-speed chemical imaging, serving as a general processing modality that can be seamlessly integrated across various spectroscopic hardware and methodologies.

Despite its superior performance, several challenges remain to be addressed. A primary limitation is the current reliance on a standardized input length, which necessitates spectral interpolation or cropping. This preprocessing may distort the local statistical distribution of noise and the fine profiles of signals, introducing subtle pixel-to-pixel correlations that can partially violate the independence assumption of single-pixel masking strategy. Furthermore, for spectra with peak intensities differing by orders of magnitude, the restoration of ultra-low-intensity peak shapes remains an unresolved problem. To mitigate this in the current stage, F2P employs a segment-wise processing strategy. Future research will focus on developing adaptive architectures and incorporating physical priors to better distingush signals from stochastic noise at these fundamental physical limits. For complex experimental scenarios not fully covered by pre-training, self-supervised fine-tuning serves as a robust supplementary strategy for recovering subtle details (see Table S8), despite the added computational cost. Supervised fine-tuning (SFT) also shows increasing benefits under stronger instrument noise added to simulated spectra (see Table S9), although its effectiveness in real-world applications requires further investigation.

# AUTHOR INFORMATION

## Corresponding Authors

**Hao Ma** − *State Key Laboratory of Physical Chemistry of Solid Surfaces, Collaborative Innovation Center of Chemistry for Energy Materials (iChEM), College of Chemistry and Chemical Engineering, Xiamen University, Xiamen 361005, P. R. China; Institute of Artificial Intelligence, Xiamen University, Xiamen 361005, China* ; *Tan Kah Kee Innovation Laboratory, Xiamen 361005, P. R. China; orcid.org/0000-0001-8341-4003;*

Email: oaham@xmu.edu.cn; Fax: +86-0592-2185889

**Guo-Kun Liu** − *State Key Laboratory of Marine Environmental Science, Fujian Provincial Key Laboratory for Coastal Ecology and Environmental Studies, Center for Marine Environmental Chemistry & Toxicology, College of the Environment and Ecology, Xiamen University, Xiamen 361102, China. orcid.org/0000-0003-2501-7178;*

Email: guokunliu @xmu.edu.cn; Fax: +86-0592-2185889

**Bin Ren** − *State Key Laboratory of Physical Chemistry of Solid Surfaces, Collaborative Innovation*

*Center of Chemistry for Energy Materials (iChEM), College of Chemistry and Chemical Engineering, Xiamen University, Xiamen 361005, P. R. China; Institute of Artificial Intelligence, Xiamen University, Xiamen 361005, China ; Tan Kah Kee Innovation Laboratory, Xiamen 361005, P. R. China; orcid.org/0000-0002-9821-5864;*

Email: bren@xmu.edu.cn; Fax: +86-0592-2185889

## Authors

**Nan-Nan Zhang** - *State Key Laboratory of Physical Chemistry of Solid Surfaces, Collaborative Innovation Center of Chemistry for Energy Materials (iChEM), College of Chemistry and Chemical Engineering, Xiamen University, Xiamen 361005, P. R. China; Institute of Artificial Intelligence, Xiamen University, Xiamen 361005, China*

**Si-Heng Luo** - *Strait Institute of Flexible Electronics (SIFE, Future Technologies), Fujian Key Laboratory of Flexible Electronics, Fujian Normal University and Strait Laboratory of Flexible Electronics (SLoFE), Fuzhou, 350117, China*

**Xin-Di Pei** - *State Key Laboratory of Physical Chemistry of Solid Surfaces, Collaborative Innovation Center of Chemistry for Energy Materials (iChEM), College of Chemistry and Chemical Engineering, Xiamen University, Xiamen 361005, P. R. China; Tan Kah Kee Innovation Laboratory, Xiamen 361005, P. R. China*

**Wen-Xuan Li** - *State Key Laboratory of Physical Chemistry of Solid Surfaces, Collaborative Innovation Center of Chemistry for Energy Materials (iChEM), College of Chemistry and Chemical Engineering, Xiamen University, Xiamen 361005, P. R. China; Tan Kah Kee Innovation Laboratory, Xiamen 361005, P. R. China*

**Jun Wang** - *State Key Laboratory of Physical Chemistry of Solid Surfaces, Collaborative Innovation Center of Chemistry for Energy Materials (iChEM), College of Chemistry and Chemical Engineering, Xiamen University, Xiamen 361005, P. R. China; Tan Kah Kee Innovation Laboratory, Xiamen 361005, P. R. China*

**Yi-Fan Bao** - *State Key Laboratory of Physical Chemistry of Solid Surfaces, Collaborative Innovation Center of Chemistry for Energy Materials (iChEM), College of Chemistry and Chemical Engineering, Xiamen University, Xiamen 361005, P. R. China; Tan Kah Kee Innovation Laboratory, Xiamen 361005, P. R. China*

### Author Contributions

All authors have given approval to the final version of the manuscript.

### Notes

The authors declare no competing financial interest.

### Data availability

The datasets generated and analyzed are available in the Hugging Face repository: https://huggingface.co/datasets/moria5161/fast2peak. The corresponding model weights are available in the Hugging Face repository: https://huggingface.co/moria5161/fast2peak.

### Code availability

The code for this work is available for open source at: https://github.com/moria5161/fast2peak.

## ACKNOWLEDGMENTS

This work was supported by the National Natural Science Foundation (Grant No: 22227802, 22595412, 22474117, and 22272139) of China and the Fundamental Research Funds for the Central Universities (20720250080).

# Supporting Information

## Principle of self-supervised restoration

The first self-supervised denoising work in the field is Noise2Noise (N2N)[1]。N2N removes the dependency on clean samples for noise reduction, demonstrating that a model can still converge to clean images by learning only a "noisy image to noisy image" mapping. The underlying rationale can be illustrated through the following derivation:

For supervised learning (SL), the essence of denoising lies in the model learning the mapping from a noisy image to a clean image. The objective function is as follows:

$$\theta_{N2C} = \arg\min_{\theta} \mathbb{E}[\| f_\theta(y_1) - x \|_2^2] \tag{1}$$

Where $\theta$ represents the parameters optimized during model training. Under the supervised learning (SL) paradigm, the denoising task is commonly referred to as Noise2Clean (N2C). Its objective function is typically defined as an L2 loss term:$\| f_\theta(y_1) - x \|_2^2$, where $y_1$ is the noisy image, $f_\theta(y_1)$ is the model's output, and $x$ is the ground-truth (clean) data. Thus, the model is optimized toward minimizing the expectation of the L2 term. In contrast, the objective function of N2N is as follows:

$$\theta_{N2N} = \arg\min_{\theta} \mathbb{E}[\| f_\theta(y_1) - y_2 \|_2^2]$$

$$= \arg\min_{\theta} \mathbb{E}[\| f_\theta(y_1) - x - e_2 \|_2^2]$$

$$= \arg\min_{\theta} \mathbb{E}[\| f_\theta(y_1) \|_2^2 - 2x^\top f_\theta(y_1) - 2e_2^\top f_\theta(y_1) + \| x \|_2^2]$$

$$= \arg\min_{\theta} \mathbb{E}[\| f_\theta(y_1) \|_2^2 - 2x^\top f_\theta(y_1) + \| x \|_2^2]$$

$$= \arg\min_{\theta} \mathbb{E}[\| f_\theta(y_1) - x \|_2^2] \tag{2}$$

Where $y_2$ is another observation corresponding to the same ground-truth $x$ as $y_1$, but with noise independent of that in $y_1$. Assuming the noise follows an independent and identically distributed (i.i.d.) distribution with zero mean, we have $y_2 = x + e_2$ and $\mathbb{E}(e_2) = 0$. Substituting this into Equation (2) and omitting the $e_2$ term, after simplification, Equation (2) reduces to Equation (1). The statistical equivalence between Equations (1) and (2) fully reveals the core principle of self-supervised denoising: a model can converge to the statistical average of the underlying signal by learning the mapping between two independent noisy observations.

Our single-pixel masking strategy was inspired by Noise2Void (N2V)[2]. For spectral data, we assume that neighboring signal values are correlated, whereas neighboring noise values are independent and uncorrelated. This assumption can be expressed probabilistically as follows:

$$y = x + n \tag{3}$$

$$p(x_i|x_j) \neq p(x_i) \tag{4}$$

$$p(n|x) = \prod_i p(n_i|x_i) \tag{5}$$

In Equation (3), $y$ represents a noisy spectrum, $x$ the corresponding clean signal spectrum, and $n$ the noise distribution of the spectrum.

In Equation (4), $x_i$ and $x_j$ are two adjacent true signal points in $x$, $p(x_i)$ denotes the probability distribution of $x_i$, and $p(x_i \mid x_j)$ denotes the conditional probability distribution of $x_i$ given the neighboring signal point $x_j$.

In Equation (5), $p(n \mid x)$ represents the conditional probability distribution of $n$ given the spectrum $x$, and $\prod_i p(n_i \mid x_i)$ represents the joint conditional probability distribution of the noise points given the signal at all indexed positions of $x$.

The original N2V uses a U-Net trained on unpaired noisy images. To achieve efficient spectral reconstruction, F2P replaces the traditional convolutional neural network (CNN) with a Transformer, thereby enhancing the ability to capture global correlations of complex spectral features. Trained on the 400,000-sample training set of F2P-B, the model learns the general principles governing spectral peak shape patterns and thus achieves universal denoising performance. Although simulated data cannot yet fully cover all real-world complex experimental spectra, we believe that fine-tuning or localized cut-based focused processing will further improve the generalization capability of F2P on experimental data.

## Datasets

To establish a universal benchmark for objective evaluation of spectral denoising algorithms, we constructed a comprehensive data library organized into three tiers of increasing experimental authenticity (F2P Benchmark, F2P-B): the F2P-Syn, F2P-Semi, and F2P-Real datasets. F2P-Syn comprises a training set of 400,000 synthetic spectra generated via pseudo-Voigt profiles superimposed with realistic Gaussian-Poisson noise and complex backgrounds, along with a test set of 1,024 simulated spectra with precise ground truth. F2P-Semi was constructed by superimposing authentic instrumental noise onto ultra-high SNR experimental spectra, which serve as ground-truth proxies acquired from 2D materials, HeLa cells, and PEO+LiTFSI electrolytes (**Collection conditions for experimental data**). This semi-synthetic strategy preserves the non-ideal characteristics of real instrumental noise while retaining precise ground truth for rigorous quantitative evaluation. F2P-Real consists of pure experimental datasets without ground truth, including PEO+LiTFSI spectra acquired across five commercial spectrometers, hyperspectral $WSe_2$ mapping at varying integration times, and real-time graphene mapping on a custom EX-Raman system.

## Method for generating simulated data

The standard Voigt lineshape is the convolution of a Gaussian function and a Lorentzian function. To improve computational efficiency, we adopt the pseudo-Voigt approximation widely used in spectroscopy, which approximates it as a linear combination of the two. The formula is as follows:

$$V(x,\mu,w) = \rho \times G(x,\mu,w) + (1-\rho) \times L(x,\mu,w) \tag{6}$$

$$G(x,\mu,w) = \frac{1}{\sigma\sqrt{2\pi}} e^{-\frac{(x-\mu)^2}{2\sigma^2}}, \sigma = \frac{w}{2\sqrt{2\ln 2}} \tag{7}$$

$$L(x,\mu,w) = \frac{1}{\pi} \cdot \frac{w/2}{(x-\mu)^2 + (w/2)^2} \tag{8}$$

Where $\mu$ is the peak position, $w$ is the full width at half maximum (FWHM), G is the Gaussian component, L is the Lorentzian component, and $\rho$ denotes the contribution of the Gaussian function to the pseudo-Voigt profile. The formula for generating a Raman spectrum with multiple peaks is as follows:

$$R = \sum_{i=1}^{N} S(i) = \sum_{i=1}^{N} A(i) \times V\big(x, U(i), W(i)\big) \tag{9}$$

Where $A$ denotes the amplitude of the signal, $U$ the peak position, $W$ the FWHM, $N$ the total number of Raman peaks. The parameters of the generator are randomly selected within specified ranges. The simulated spectrum has a fixed length of 1600, with $\rho$=0.6785, $N$∈[5, 10], $U$∈[300, 1300], $A$∈[100, 550], $W$∈[5, 130]。

After generating the clean spectrum based on the pseudo-Voigt function, we add Gaussian-Poisson mixed random noise to the simulated spectrum to obtain the simulated noisy spectrum. The formula for the mixed noise is as follows:

(1) Gaussian noise:

$$x_{\text{noisy}} = x + \sigma \cdot \epsilon\big(\epsilon \sim \mathcal{N}(0,1)\big) \tag{10}$$

(2) Poisson noise:

$$x_{\text{noisy}} = \frac{1}{k}\mathcal{P}(x \cdot k) \tag{11}$$

Where $x$ represents the simulated spectrum. In Equation (10), $\epsilon$ is a random variable following the standard normal distribution $\mathcal{N}(0,1)$, and $\sigma$ denotes the standard deviation of the Gaussian noise, with $\sigma \in [1 \times 10^{-4}, 1 \times 10^{-1}]$. In Equation (11), $\mathcal{P}(\cdot)$ represents the Poisson distribution function, and $k$ is the photon-counting scale factor, with $k \in [2 \times 10^4, 3 \times 10^4]$.

Finally, fluorescence background and Rayleigh background are randomly added to the simulated noisy spectrum to better approximate real experimental spectral types. The mathematical models for these two backgrounds[3,4] are as follows:

(1) Fluorescence background:

$$B_{\text{Fluorescence}} = A_{\text{fluo}} \cdot \exp\left(-0.5 \cdot \left(\frac{x - x_c}{\sigma}\right)^2\right) \tag{12}$$

(2) Rayleigh background:

$$B_{\text{Rayleigh}} = A_{\text{ray}} \cdot \frac{\Gamma^2}{x^2 + \Gamma^2} \tag{13}$$

Where $x$ represents the wavenumber. In Equation (12), $A_{\text{fluo}} \in [0.1, 1.0]$ is the fluorescence peak intensity, and $x_c \in [100, 900]$ is the center position of the fluorescence. Because $x_c$ is randomly distributed at the center or edges of the spectrum, different background shapes such as a central bulge or a monotonic increase/decrease can be generated. $\sigma \in [400, 600]$ is the standard deviation of the fluorescence, where a larger $\sigma$ typically results in a broader envelope, consistent with the wide-band fluorescence background characteristics. In Equation (13), $A_{\text{ray}} \in [5 \times 10^4, 2 \times 10^5]$. Since Rayleigh scattering intensity is extremely high at low wavenumbers, this parameter is set to a large initial amplitude. $\Gamma \in [30, 60]$ controls the decay rate of the Lorentzian tail.

## Collection conditions for experimental data

To rigorously evaluate the quantitative fidelity of the F2P framework, semi-synthetic benchmark datasets with reliable ground truths were constructed following a stringent protocol inspired by He et al[5]. The workflow is as follows: (1) Ground Truth Acquisition: For each sample type, Raman spectra were acquired with prolonged integration times. To further eliminate stochastic fluctuations, 50 parallel acquisitions were performed and subsequently averaged. These ultra-high signal-to-noise ratio (SNR) cumulative spectra were designated as the ground-truth profiles. (2) Noise Extraction: Pure instrumental background noise was independently captured from the respective spectrometers under identical dark conditions. (3) Synthesis: The authentic instrumental noise components were randomly sampled and superimposed onto the corresponding ground-truth profiles, yielding highly realistic noisy spectra that faithfully emulate actual physical detection limits. The detailed sample compositions, instrumental configurations, and dataset scales utilized for constructing the semi-synthetic benchmarks are summarized in Table S1.

To thoroughly evaluate F2P's performance in transient electrochemical SERS, advanced spatiotemporal resolution and live-monitoring scenarios, specific experimental configurations were tailored for Nile Blue and 2D materials ($WSe_2$ and Graphene). (1) The working electrode was prepared by drop-depositing Ag nanoparticles onto an Ag electrode. The Ag nanoparticle-modified electrode was immersed in a $1 \times 10^{-5}$ M Nile Blue solution for 30 min, then removed and rinsed with ultrapure water. This concentration refers to the adsorption solution. Measurements were performed in a three-electrode configuration using the modified Ag working electrode and an Ag/AgCl reference electrode containing saturated KCl as its filling solution. The supporting electrolyte was 0.5 M $H_2SO_4$. (2) Hyperspectral Imaging of $WSe_2$: The spatiotemporal reconstruction limits were assessed using monolayer (1L) and bilayer (2L) $WSe_2$ flakes. Hyperspectral Raman mapping was conducted on a LabRAM HR Evolution system (Horiba) equipped with a 532 nm laser. (3) Real-Time "Acquire-and-Reconstruct" Imaging of Graphene: To demonstrate the live processing potential of the F2P framework, real-time Raman mapping of graphene was performed on a custom-built EX-Raman system (JIAYI) equipped with a 532 nm laser. The specific instrumental parameters and mapping configurations for the advanced application scenarios are detailed in Table S2.

## Evaluation metrics

The calculation methods for the RMSE, SNR, and pSNR (peak-specific SNR) metrics for spectra are as follows:

$$RMSE = \sqrt{\frac{1}{n}\sum_{i=1}^{n} (f_\theta(y_i) - x_i)^2} \tag{14}$$

$$SNR = 10 \cdot \log_{10}\left(\frac{\sum_{i=1}^{n} f_\theta(y_i)^2}{\sum_{i=1}^{n}(f_\theta(y_i) - x_i)^2}\right) \tag{15}$$

$$pSNR = \frac{I_{signal} - \mu_{noise}}{\sigma_{noise}} \tag{16}$$

Where $i$ represents the index of the Raman shift, $n$ denotes the length of the spectrum, $f_\theta(y_i)$ and $x_i$ represent the denoised value and the ground-truth value at the corresponding index, respectively. $I_{signal}$ is the intensity at the selected peak position; $\mu_{noise}$ and $\sigma_{noise}$ are the mean and standard deviation, respectively, calculated over a noise interval (i.e., a flat, signal-free region of the spectrum).

The Pearson correlation coefficient (r) was used to evaluate the overall spectral shape similarity between the restored spectrum and the reference spectrum.

$$r = \frac{\sum_{i=1}^{L}(x_i - \bar{x})\left(\hat{x}_i - \bar{\hat{x}}\right)}{\sqrt{\sum_{i=1}^{L}(x_i - \bar{x})^2}\sqrt{\sum_{i=1}^{L}\left(\hat{x}_i - \bar{\hat{x}}\right)^2}} \tag{17}$$

Here, $\bar{x}$ and $\bar{\hat{x}}$ denote the mean intensities of the reference and denoised spectra, respectively. The coefficient r ranges from −1 to 1, with values closer to 1 indicating higher linear agreement between the spectral profiles. However, this metric alone cannot reflect errors arising from overall intensity scaling or baseline offsets.

Peak F1 is used to evaluate the detection and preservation of spectral peaks after denoising. It is defined as the harmonic mean of peak detection precision and recall:

$$\text{Precision} = \frac{TP}{TP + FP}, \qquad \text{Recall} = \frac{TP}{TP + FN} \tag{18}$$

$$\text{Peak F1} = \frac{2\ \text{Precision Recall}}{\text{Precision} + \text{Recall}} = \frac{2TP}{2TP + FP + FN} \tag{19}$$

Where, TP (True Positives): Number of predicted peaks successfully matched to reference peaks. FP (False Positives): Number of predicted peaks that do not match any reference peak, representing false detections. FN (False Negatives): Number of reference peaks not matched by any predicted peak, representing missed detections.

The formulas for the image-based metrics PSNR and SSIM are as follows:

$$PSNR(x, y) = 10 \cdot log_{10}\left(\frac{MAX_I^2}{\mathrm{MSE}(x, y)}\right) \tag{20}$$

Where $x$ is the ground-truth (GT) image, $y$ is the denoised image, $MAX_I$ is the maximum pixel value of the image, and MSE is the mean squared error (the square of Equation (14)).

$$SSIM(x, y) = \frac{(2\mu_x\mu_y + C_1)(2\sigma_{xy} + C_2)}{(\mu_x^2 + \mu_y^2 + C_1)(\sigma_x^2 + \sigma_y^2 + C_2)} \tag{21}$$

Where $\mu_x$ and $\mu_y$ are the pixel-wise means of images $x$ and $y$, $\sigma_x^2$ and $\sigma_y^2$ are their variances, and $\sigma_{xy}$ is their covariance. $c_1$ and $c_2$ are stabilization constants to prevent division by zero.

The spatial resolution was quantified from an intensity profile crossing the material boundary. The profile was mildly smoothed with a Gaussian filter to suppress high-frequency fluctuations, and its first derivative was then calculated to obtain the line spread function (LSF):

$$L(x) = \frac{d}{dx}\left[G_{\sigma_s}(x) * I(x)\right] \tag{22}$$

The resulting LSF was fitted with a Gaussian function, and the spatial resolution was defined as the full width at half maximum (FWHM) of the fitted profile:

$$R_{\text{spatial}} = \text{FWHM} = 2\sqrt{2\ln 2}\,\sigma \approx 2.355\sigma \tag{23}$$

Here, $R_{\text{spatial}}$has the same spatial unit as $x$, for example $\mu$m. A smaller FWHM indicates a narrower edge response and therefore a higher spatial resolution.

## Asynchronous end-to-cloud pipeline

We integrated F2P into a self-built Raman spectrometer and successfully constructed a real-time online denoising system based on an asynchronous end-to-cloud processing mechanism. This system primarily consists of two synergistic components: hardware temporal decoupling and cloud-based ultra-fast inference.

In the first part, this study designed a synergistic asynchronous end-to-cloud architecture. This architecture leverages the hardware motion gaps inherent in point-scanning imaging using a Raman microscope (i.e., the time required for the stage to move from position $t$ to $t+1$) to establish an asynchronous communication mechanism between the instrument side and the computing side (e.g., cloud server or local workstation). The system operation is governed by the following strict temporal constraint equation:

$$\Delta T_{Trans} + \Delta T_{process} \leq \Delta T_{move} + \Delta T_{exposure} \tag{24}$$

Where $\Delta T_{\mathrm{Trans}}$ is the round-trip transmission time between the device and the cloud, $\Delta T_{\mathrm{process}}$ is the ultra-fast forward inference time of the model in the cloud, $\Delta T_{\mathrm{move}}$ is the mechanical movement time of the stage to the next pixel, and $\Delta T_{\mathrm{exposure}}$ is the spectral acquisition (exposure) time for the next pixel. When the device completes spectral acquisition at position $t+1$, the spectrum at position $t$ has already been denoised and returned to the device side.

In the second part, to meet the requirement for ultra-fast inference $\Delta T_{\mathrm{process}}$ under the above temporal constraints, a deep learning algorithm (in this study, the F2P algorithm, which achieved state-of-the-art results in benchmark evaluations) is deployed to the cloud for online inference. Specifically, the pre-trained F2P model is pre-loaded into the cloud to maximize the cloud-side processing speed.

## Supplementary Figures and Tables

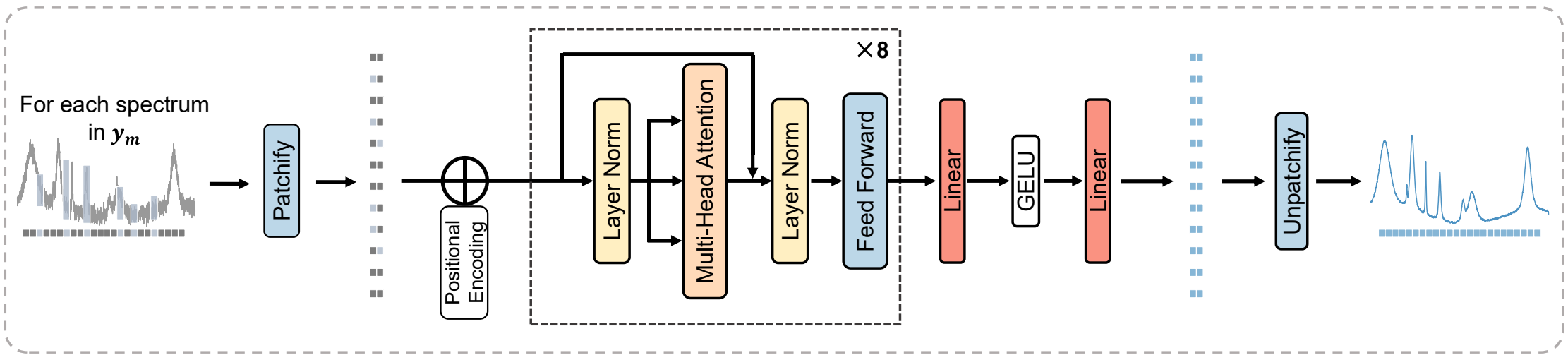

**Figure S1.** The overall architecture of the Spectral Transformer.

The overall architecture of the Spectral Transformer is shown in Figure S1. It adopts a Transformer encoder-based architecture for end-to-end reconstruction of one-dimensional spectra standardized to a length of 1600. Given a masked spectrum $y_m$ with dimensions B × 1 × 1600, a one-dimensional convolutional Patch Embedding layer first performs patch partitioning and feature embedding. Both the kernel size and stride are set to 16, with an embedding dimension of 256, converting each spectrum into 100 non-overlapping patches and yielding a token sequence of dimensions B × 100 × 256. This convolutional operation integrates patchification and feature embedding without requiring an additional embedding layer. The resulting token sequence is then processed by a backbone consisting of eight stacked RPE Transformer encoder blocks. Each block follows a Pre-Norm design and sequentially comprises Layer Normalization, multi-head relative-position self-attention, a residual connection, a second Layer Normalization, a feed-forward network, and another residual connection. The multi-head attention module contains eight attention heads, corresponding to a feature dimension of 32 per head, while the query, key, and value representations are jointly projected from 256 to 768 dimensions using bias-free linear layers. Learnable relative positional information is incorporated into the attention mechanism to capture both local and long-range dependencies along the spectral sequence. No attention dropout, MLP dropout, or stochastic depth is applied. The feed-forward network in each Transformer block uses a fourfold expansion with a 256 → 1024 → 256 structure and GELU activation. After feature extraction through the eight encoder blocks, each token is mapped to 16 spectral points through a lightweight prediction head with a 256 → 512 → 16 architecture, again using GELU activation in the hidden layer. The 100 predicted patches are subsequently rearranged in their original order through an unpatchify operation to recover the complete output spectrum with dimensions B × 1 × 1600. The model contains approximately 6.47 million parameters.

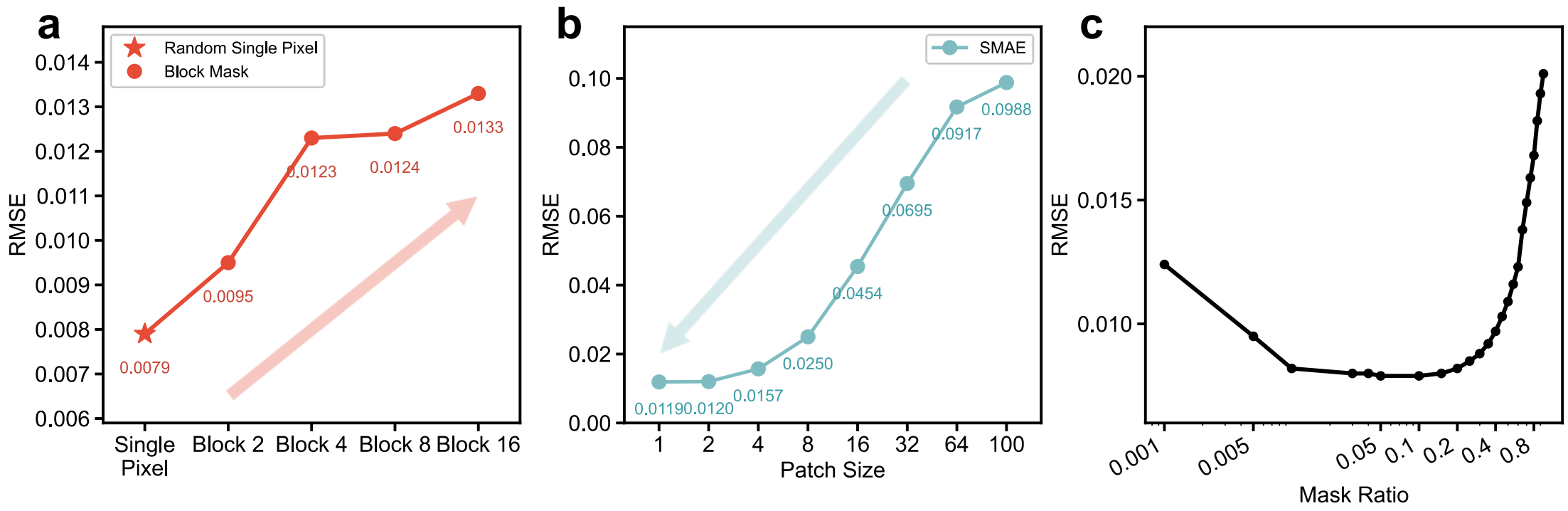

**Figure S2.** F2P ablation experiment results. (a) restoration RMSE trend of F2P on the Simu test set using block-wise corruption masks of varying sizes. (block corruption is similar to the Gaussian peak masking used in P2P) (b) restoration RMSE trend of SMAE on the Simu test set using different patch sizes.(c) The influence of mask ratio on noise reduction for 1024 simulated noise spectra.

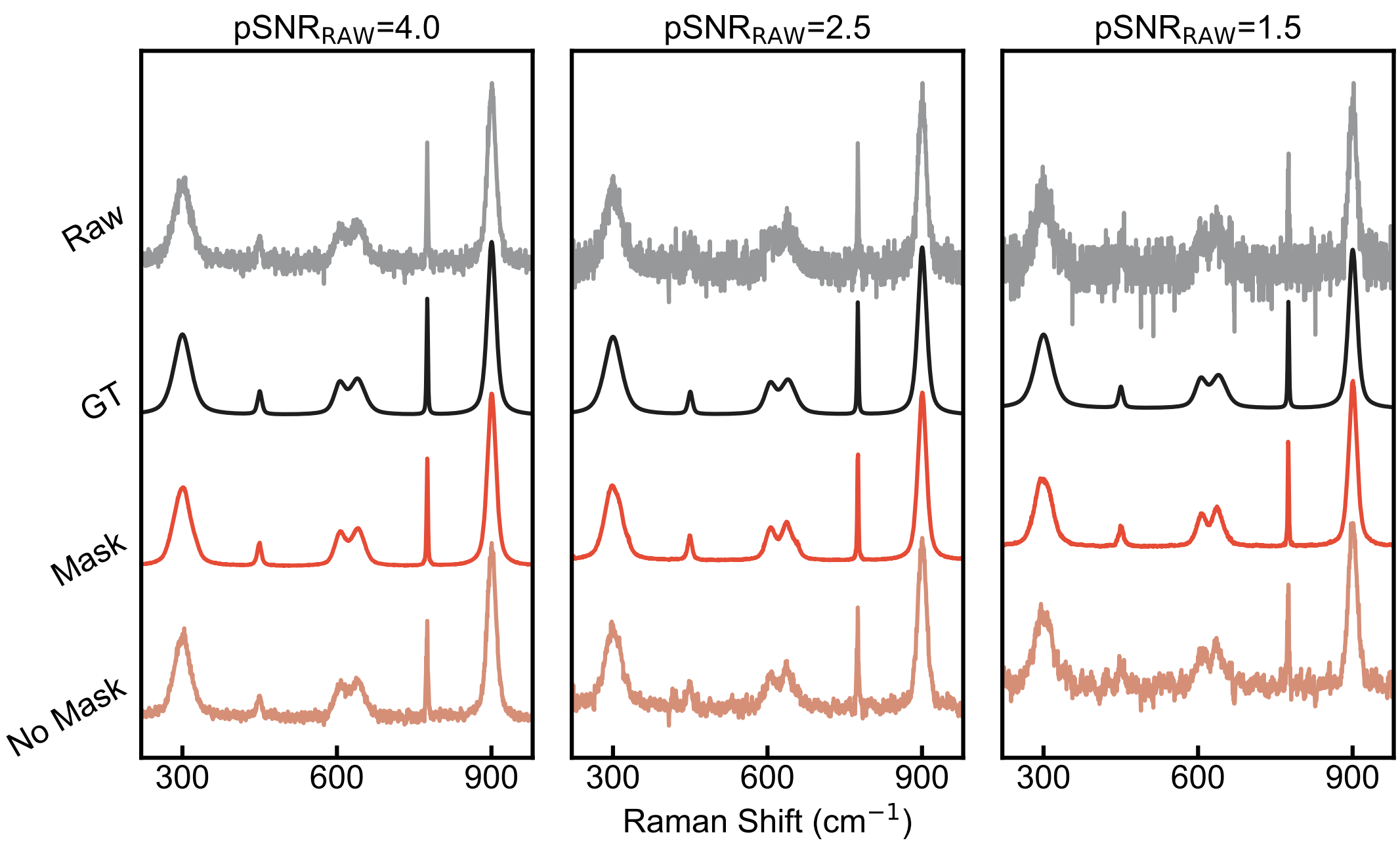

**Figure S3.** Comparison of noise reduction performance between masked and unmasked training models.

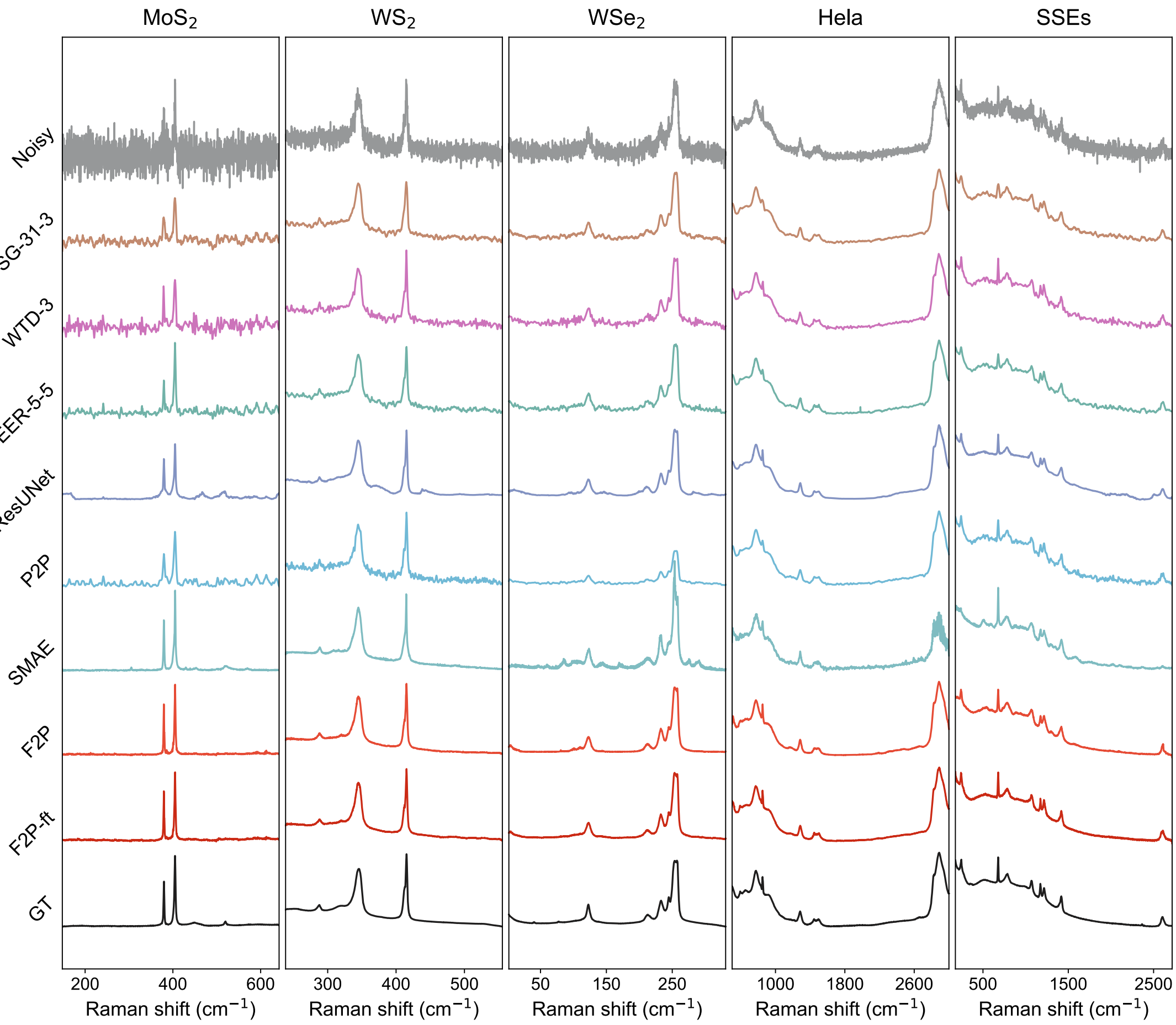

**Figure S4.** Comparison chart of denoising results for the 2D Materials, Hela, and SSEs datasets, with one sample selected from each set as an example.

As shown in Figure S4, F2P was applied to datasets from diverse experimental systems, including two-dimensional materials, solid-state electrolytes (SSEs), and biological cells, achieving effective noise suppression across all datasets. Although F2P showed certain limitations on the SSEs dataset, particularly in preserving the intensity of the δ(SNS) peak at 740 $cm^{-1}$, its performance was further improved through self-supervised fine-tuning using the experimental data themselves (F2P-ft, Figure S4 and Table S9), without requiring clean ground-truth spectra. This improvement further demonstrates the adaptability of F2P to previously unseen experimental data. Fine-tuning was performed for 100 epochs, with the learning rate gradually decreasing from $1\times10^{-5}$ to $1\times10^{-6}$.

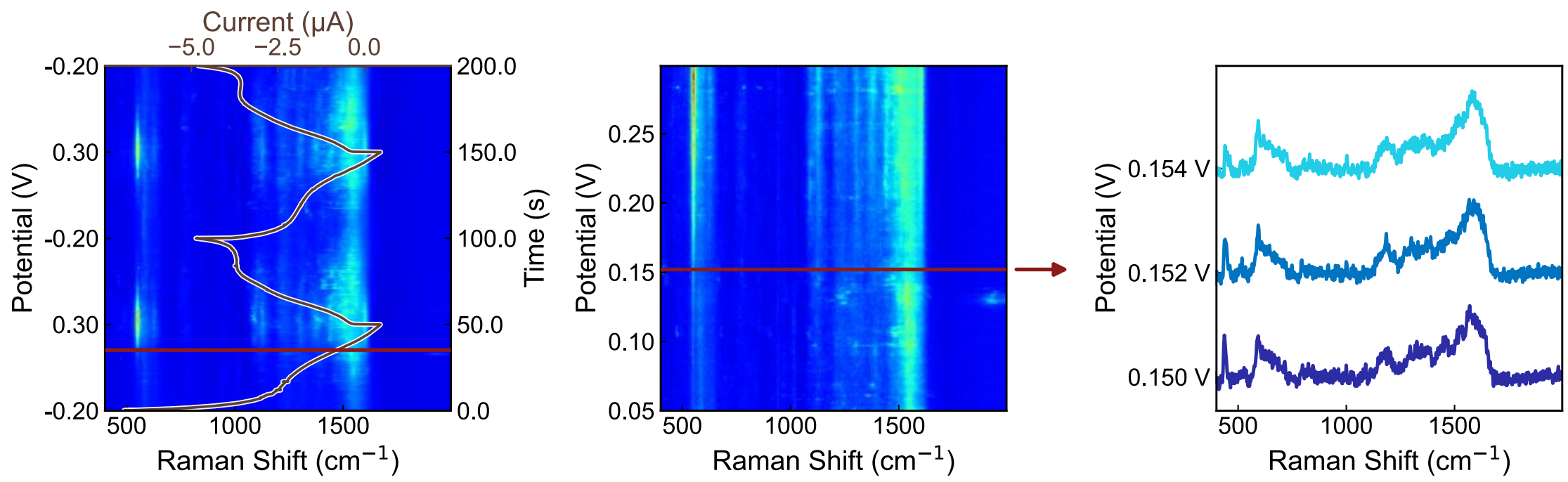

**Figure S5.** Reference results for Nile Blue EC-SERS at 100 ms integration time.

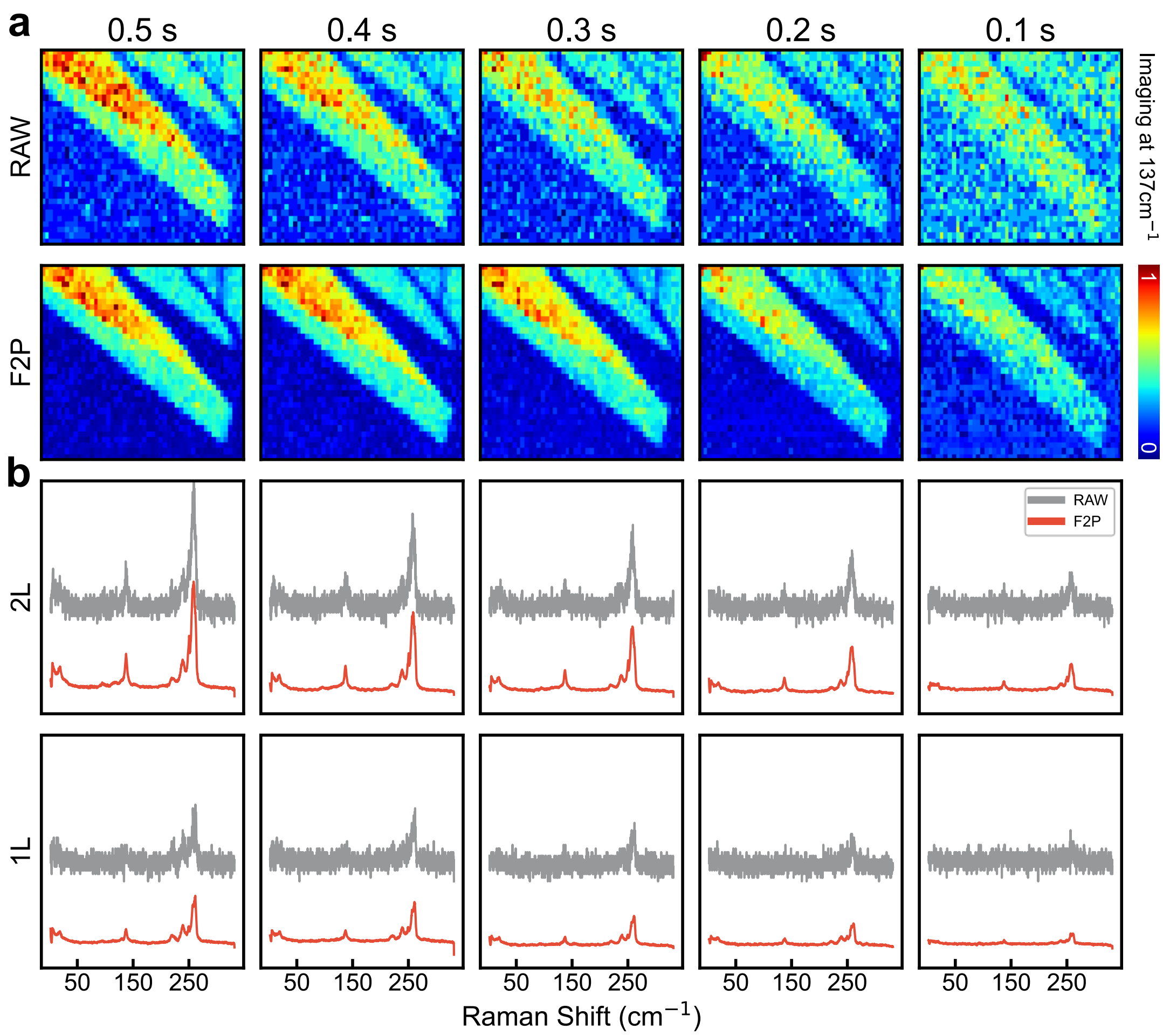

**Figure S6.** F2P is applied to $WSe_2$ rapid imaging denoising. (a) Peak intensity imaging of the original and F2P at the 137 $cm^{-1}$ for different integration times. (b) Original spectra and denoised spectra at selected positions of 1L and 2L under different times.

Rapid Raman hyperspectral imaging yields crucial insights into the spatial heterogeneities of complex material systems, which are essential for understanding structure-property relationships. Taking two-dimensional (2D) materials as an example, we demonstrate restoration performance of F2P using Raman intensity mapping of $E_{1g}^2$ peak of $WSe_2$ at 137 $cm^{-1}$ across varying integration times (0.5 s to 0.1 s), as shown in Figure S6a. To quantitatively assess restoration fidelity, the peak area imaging of $A_{1g}$ peak at 255-270 $cm^{-1}$ for 0.5 s was utilized as the proxy GT (Figure S7). Additionally, Figure S6b extracts raw and denoised spectra from specific monolayer (1L) and bilayer (2L) sites (marked with black star).

The clear boundaries demonstrate that F2P significantly enhances both global clarity and local fidelity. Specifically, at 0.5 s, F2P improved the PSNR from 19.64 dB to 27.73 dB and surged the SSIM by 45% (Table S10). At the microscopic level, the faint 137 $cm^{-1}$ characteristic peaks on both 1L and 2L sites were precisely restored. When integration time was reduced to 0.3 s, fine contours in the raw imaging became severely blurred. Nonetheless, F2P successfully reconstructed clear morphological boundaries. At the extreme 0.1 s limit, F2P maintained a 67% relative advantage in SSIM (0.4017 vs. 0.2406). However, the PSNR gain was notably constrained to only 2.68 dB. As the spectral comparison reveals (Figure S6b), such severe noise completely overwhelms the effective photon information of the 137 $cm^{-1}$ peak, rendering feature extraction nearly impossible.

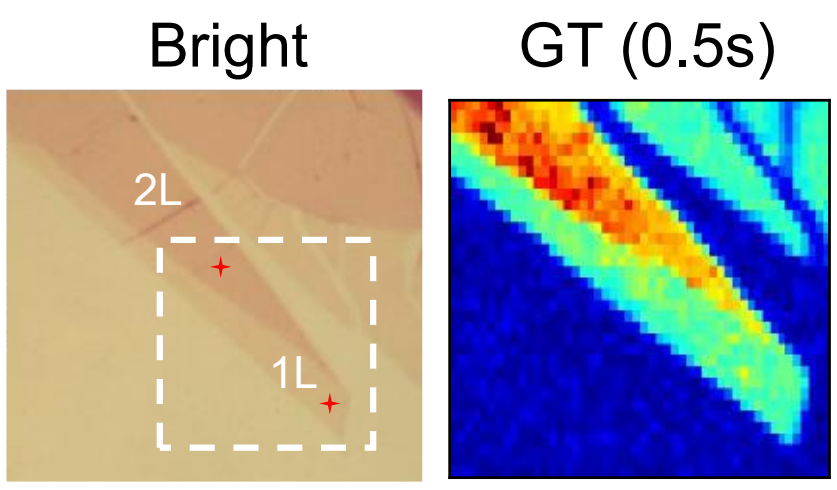

**Figure S7.** White light imaging of $WSe_2$ and peak area imaging (as GT) at 255 $cm^{-1}$ under 0.5s integration.

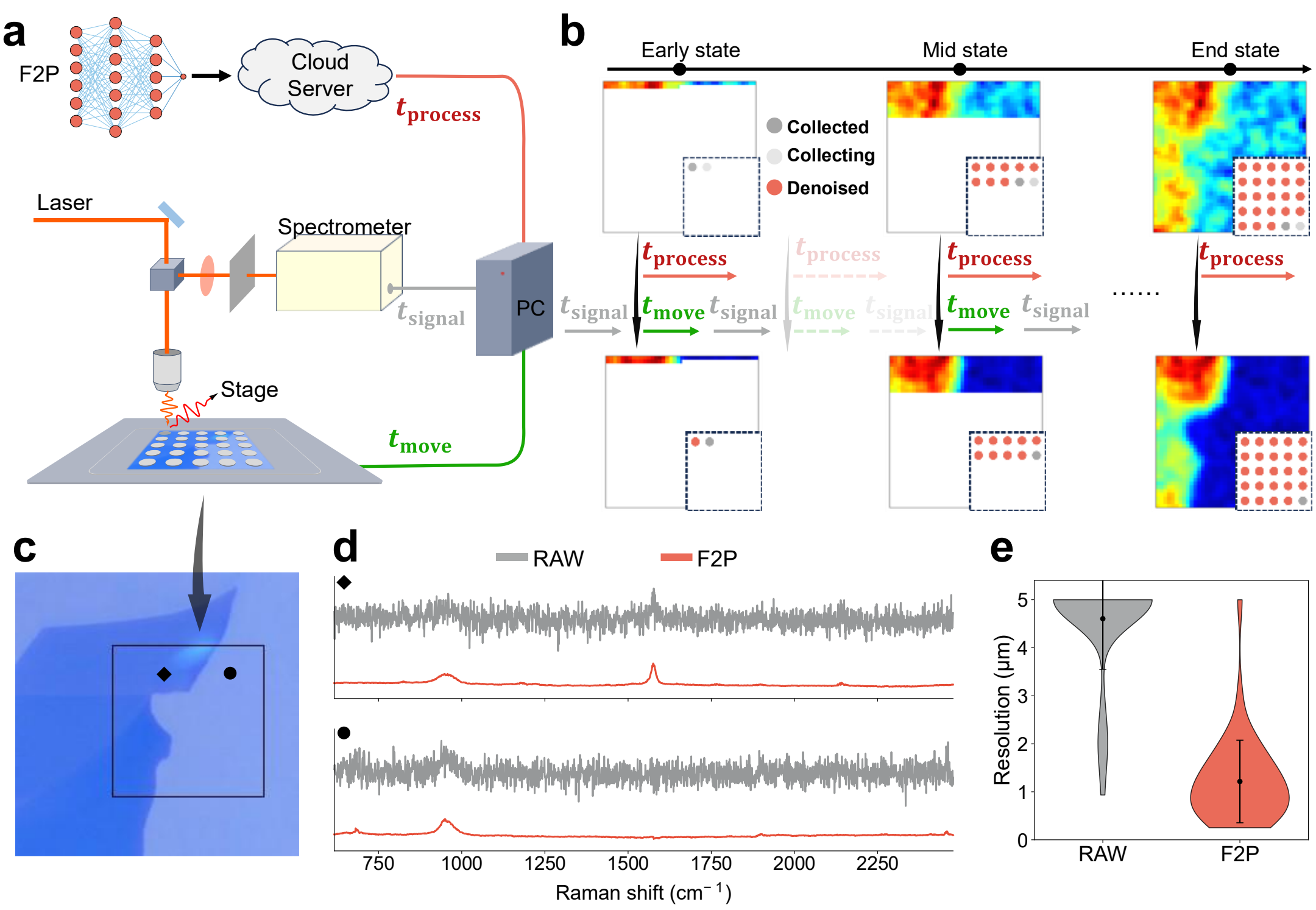

**Figure S8.** Real-time F2P-enabled hyperspectral Raman imaging. (a) Asynchronous end-to-cloud system for spectral acquisition and F2P restoration. (b) Progressive reconstruction of raw and F2P-restored Raman images during raster scanning, with inset schematics showing the corresponding asynchronous processing states. (c) Bright-field image of graphene showing the Raman mapping region and two representative positions. (d) Raw and F2P-restored Raman spectra at the two positions marked in (c). (e) Spatial-resolution distributions of the raw and F2P-restored Raman images.

Applying F2P to local instruments still requires high-performance computational resources that may not be available on many Raman instruments and other spectroscopic platforms. For rapid hyperspectral chemical mapping, this constraint is particularly acute, as the bottleneck shifts from temporal to computational throughput. We therefore implemented an asynchronous end-to-cloud pipeline. As shown in Figure S8a, F2P is deployed on the cloud through a Flask-based service and communicates with the local hardware through the PC. During raster scanning, each acquired spectrum is transmitted to the cloud for F2P restoration and subsequently returned to the instrument side. The lower-right corner of each subpanel in Figure S8b shows the corresponding theoretical schematic of the asynchronous processing workflow. In this workflow, once the spectrum at the current pixel (dark-gray dot) is transmitted to the cloud, the stage immediately moves to the next position and starts the next acquisition (light-gray dot), while F2P restoration proceeds in parallel (red dot when processed). The processing time $t_{\text{process}}$, which consists of the F2P inference time and the data communication time, is fully masked by the mechanical dwell time ($t_{\text{move}} + t_{\text{signal}}$), allowing restoration of the previous spectrum to proceed in parallel with acquisition of the next one. Through this asynchronous pipeline, the system achieves a seamless temporal overlap between hardware data collection and algorithmic inference, ultimately realizing a genuine real-time "acquire-and-process" performance, where spectral restoration is completed in parallel with hardware data collection.

Taking graphene at 1580 $cm^{-1}$ ($E_{2g}$) as an example (the bright-field image is shown in Figure S8c), the total processing time (~40 ms for F2P inference time and ~100 ms for data communication time) was fully masked by the mechanical dwell time. As shown by the experimental images in Figure S8b, the Raman image was progressively reconstructed during scanning: severe fluctuations in the raw data initially obscured the graphene boundary, whereas the F2P-restored image progressively recovered the spatial contrast and was completed by the end of the scan (see Videos S1 and S2 for details). Spectra acquired at the two representative positions marked in Figure S8c further confirm that F2P substantially suppresses stochastic fluctuations while preserving the characteristic graphene Raman features (Figure S8d). Importantly, the restored Raman image (the complete Raman image in Figure S8b) closely reproduces the graphene morphology observed in the bright-field micrograph. Quantitatively, the effective edge-localization width was reduced from 4.57 to 1.24 μm, corresponding to a 3.7-fold improvement in spatial resolution (Figure S8e; details of the calculation are provided in the Supporting Information). This improvement is particularly notable under the 50 ms acquisition condition, where the reduced photon budget would otherwise compromise the spatial fidelity of the raw measurement (Table S7). Unlike conventional "acquire-then-process" workflows that impose a "dark period" of offline restoration, F2P enables "see-as-you-scan" autonomy, establishing a robust algorithmic foundation for future AI-driven, high-throughput, and autonomous material characterization.

**Table S1.** Detailed acquisition parameters for the semi-synthetic benchmark datasets.

| Sample type | Materials | Spectrometer | Laser wavelength | Data pairs |
|---|---|---|---|---|
| 2D Materials (2DMs) | $MoS_2$, $WS_2$, $WSe_2$ | LabRAM HR-Evolution (Horiba) | 532 nm | 72 |
| Biological Cells (Hela) | Hela cells | Raman-11 (Nanophoton) | 532 nm | 32 |
| Solid-State Electrolytes (SSEs) | PEO+LiTFSI | Raman-11 (Nanophoton) | 532 nm | 32 |

**Table S2.** Acquisition configurations for hyperspectral and real-time Raman imaging.

| Application Scenario | Sample type | Spectrometer | Laser wavelength |
|---|---|---|---|
| Spatiotemporal Limit Evaluation | 1L & 2L $WSe_2$ | LabRAM HR-Evolution (Horiba) | 532 nm |
| TEC-SERS | Nile Blue | EC-Raman-L215 (Jayitec) | 633 nm |
| Live "Acquire-and-Reconstruct" | Graphene | EX-Raman (Jayitec) | 532 nm |

**Table S3.** The influence of γ on the performance of F2P.

| γ | 3 | 7 | 11 | 15 | 19 |
|---|---|---|---|---|---|
| RMSE | 0.0081 | 0.0079 | 0.0080 | 0.0079 | 0.0080 |

**Table S4. Quantitative comparison on the Maybridge $^1$H NMR denoising benchmark.**

| Methods | Pearson ↑ | RMSE (×10⁻³) ↓ | SNR (dB) ↑ | Peak F1 ↑ |
|---|---|---|---|---|
| RAW | 0.692937 | 53.313 | 2.187 | 0.039786 |
| WT | 0.912430 | 16.468 | 11.405 | 0.349464 |
| Cadzow | 0.961960 | 13.464 | 12.295 | 0.489976 |
| LD-Net | 0.973669 | 8.418 | 16.966 | 0.555886 |
| **DN-Unet** | **0.986677** | **6.331** | **19.347** | **0.831536** |
| F2P | 0.986635 | 6.703 | 18.311 | 0.782349 |

The NMR benchmark was constructed from the publicly available Maybridge $^1$H NMR dataset reported by Castillo et al[6], comprising 1001 experimental spectra. Gaussian noise with σ = 0.01, 0.05, and 0.10 was introduced to construct the test data. Compared with the noisy input, conventional WT and Cadzow filtering[7], and the supervised DN-Unet[8] and LD-Net[9] models, F2P achieved a Pearson correlation of 0.9866, an RMSE of $6.703 \times 10^{-3}$, an SNR of 18.31 dB, and a Peak F1 of 0.782. Without dataset-specific retraining or fine-tuning, F2P substantially outperformed the conventional methods and LD-Net, while achieving performance comparable to the supervised DN-Unet, demonstrating strong zero-shot transferability to experimental NMR spectra.

**Table S5. Performance comparison on the reconstructed FTIR denoising benchmark.**

| Methods | Pearson mean ↑ | Pearson median ↑ | RMSE ↓ | SNR (dB) ↑ |
|---|---|---|---|---|
| RAW | 0.8503 | 0.8726 | 0.0748 | 5.57 |
| SavGol | 0.9143 | 0.9316 | 0.0457 | 10.04 |
| Autoencoder | 0.9444 | 0.9462 | 0.0408 | 11.55 |
| **F2P** | **0.9849** | **0.9869** | **0.0217** | **16.38** |

To further evaluate the generalization capability of F2P on external FTIR data, we reconstructed a pure denoising benchmark based on the publicly available FTIR dataset of Brandt et al[10]. The original study considered both random noise and baseline distortions. To align the evaluation task with the noise-suppression objective of F2P, additional baseline perturbations were excluded, and only noise-corrupted spectra paired with their clean references were constructed, resulting in a test set of 4,000 spectra. As a supervised learning baseline, the shallow Autoencoder proposed in the original study was retrained on the reconstructed dataset using the same training-test setting, whereas F2P was applied directly for inference without any retraining or fine-tuning on the FTIR data. F2P achieved the best performance across all core metrics, with Pearson mean and median correlations of 0.9849 and 0.9869, respectively, compared with 0.9444 and 0.9462 for the retrained Autoencoder. Meanwhile, RMSE decreased from 0.0408 to 0.0217, and SNR increased from 11.55 dB to 16.38 dB. These results demonstrate that F2P can effectively transfer to a new spectroscopic modality while maintaining high spectral reconstruction fidelity, even without FTIR-specific retraining.

**Table S6. Quantitative comparison on the experimental PXRD denoising benchmark.**

| Methods | Pearson ↑ | RMSE ↓ | SNR (dB) ↑ | Peak F1 ↑ |
|---|---|---|---|---|
| RAW | 0.8766 | 0.0455 | 9.11 | 0.264 |
| SavGol | 0.9375 | 0.0290 | 12.08 | 0.521 |
| Wavelet | 0.9472 | 0.0274 | 12.75 | 0.469 |
| **F2P** | **0.9592** | **0.0246** | **13.69** | **0.563** |

The PXRD benchmark was constructed from the publicly available experimental dataset[11]. After quality screening, 715 experimental PXRD patterns were retained as reference spectra. To evaluate robustness under diverse noise statistics, Gaussian, Poisson, and Poisson-Gaussian mixed noise were introduced at multiple noise levels. Savitzky-Golay filtering, Coiflet-4 wavelet denoising, and F2P were directly evaluated on the same test set, with F2P applied without dataset-specific retraining or fine-tuning. Averaged across all seven noise conditions, F2P achieved the highest Pearson correlation (0.9592), SNR (13.69 dB), and Peak F1 score (0.563), together with the lowest RMSE (0.0246). These results demonstrate that F2P consistently outperforms conventional filtering methods across different noise regimes while preserving the characteristic diffraction-peak structure.

**Table S7. Real-time imaging resolution of graphene at different integration times.**

| Integration Time (ms) | Raw resolution (µm) | Resolution after F2P processing (µm) |
|---|---|---|
| 100 (Reference) | 1.45 ± 0.61 | - |
| 50 | 4.57 ± 1.07 | 1.24 ± 0.84 |
| 20 | 4.70 ± 1.07 | 2.42 ± 1.31 |
| 10 | 5.00 ± 0.00 | 3.80 ± 1.96 |

Real-time Raman imaging experiments were performed at integration times of 50, 20, and 10 ms, with a 100 ms acquisition serving as the reference. Spatial resolution was evaluated row by row for each 41 × 41-pixel image using the same edge-response analysis described above, with a physical pixel spacing of 0.25 µm. Under severe noise conditions, Gaussian fitting of the line spread function failed for some rows because a reliable bell-shaped response could not be resolved, whereas valid fits could often be recovered after F2P restoration. To enable consistent quantitative comparison across all rows and imaging conditions, rows without a valid Gaussian fit were assigned a spatial resolution of 5 µm as a predefined upper-bound value, representing a severely broadened edge response with effectively unresolved spatial information. The reported values represent the mean ± standard deviation across all 41 rows, including those assigned the predefined value.

**Table S8. Evaluation results of F2P on the experiment dataset.**

| Metrics | SNR(dB) / RMSE | | |
|---|---|---|---|
| Datasets/Methods | 2DMs | Hela | SSEs |
| SG-31-3 | 18.10 / 0.0133 | 32.14 / 0.0071 | 30.73 / 0.0099 |
| WTD-3 | 18.84 / 0.0130 | 32.36 / 0.0077 | **32.30 / 0.0086** |
| PEER | 19.88 / 0.0107 | **32.71 / 0.0066** | 31.87 / 0.0088 |
| ResUNet | 19.66 / 0.0111 | 29.60 / 0.0095 | 31.66 / 0.0091 |
| P2P | 19.58 / 0.0117 | 31.54 / 0.0076 | 31.38 / 0.0098 |
| SMAE | 15.37 / 0.0177 | 26.51 / 0.0144 | 24.76 / 0.0195 |
| F2P | **23.68 / 0.0065** | 32.24 / 0.0069 | 31.74 / 0.0087 |
| **F2P-ft** | **25.62 / 0.0053** | **35.58/ 0.0047** | **36.88 / 0.0050** |

**Table S9. Performance comparison on simulated reference spectra corrupted with measured instrument noise.**

| Methods | σ ~ U(0.05, 0.10) | | σ ~ U(0.15, 0.20) | | σ ~ U(0.25, 0.30) | |
|---|---|---|---|---|---|---|
| | SNR (dB) ↑ | RMSE | SNR (dB) ↑ | RMSE | SNR (dB) ↑ | RMSE |
| Raw | 12.402 | 0.0754 | 4.919 | 0.1756 | 0.980 | 0.2757 |
| F2P | 28.932 | 0.0116 | 21.215 | 0.0277 | 17.380 | 0.0429 |
| **F2P-SFT** | **29.482** | **0.0109** | **23.001** | **0.0228** | **19.968** | **0.0327** |

Clean reference spectra were sampled from the original F2P pretraining splits: 9,000 from the training set and 1,000 from the validation set. The test set uses the 1,024 clean references from the original Simu set, but replaces its original noise with measured instrument noise scaled by σ sampled uniformly from each specified range. For each range, the same pretrained F2P was independently fine-tuned across all parameters using full-spectrum MSE without masking. Training used AdamW, a batch size of 64, weight decay of 0.001, and cosine learning-rate decay from $10^{-5}$ to $10^{-6}$ over a maximum of 100 epochs; checkpoints were selected by validation MSE. Instrument-noise-specific SFT reduced RMSE by 6.02%, 17.69%, and 23.86% as noise increased, suggesting greater benefits under stronger contamination. Future work will explore instrument-specific adaptation across diverse spectrometers and acquisition settings, with validation on real experimental spectra to assess its robustness and practical utility.

**Table S10. Comparison of Spatial Consistency between $WSe_2$ Weak-Peak Raman Images at Different Integration Times and Strong-Peak Reference Image.**

| | RAW | | F2P | |
|---|---|---|---|---|
| Integration time (s) | PSNR (dB) ↑ | SSIM ↑ | PSNR (dB) ↑ | SSIM ↑ |
| 0.5 | 19.6426 | 0.6638 | 27.7317 | 0.9629 |
| 0.4 | 18.6127 | 0.5864 | 24.2854 | 0.7757 |
| 0.3 | 16.6383 | 0.4615 | 19.6132 | 0.6356 |
| 0.2 | 15.0131 | 0.3100 | 16.5687 | 0.4820 |
| 0.1 | 12.7496 | 0.2406 | 15.4285 | 0.4025 |